# Development of a Context Aware Virtual Smart Home Simulator

Tam Van Nguyen, *Student Member, IEEE*, Huy Anh Nguyen and Deokjai Choi
*Department of Computer Engineering, Chonnam National University*
*300 Yongbong-dong, Bukgu, Kwangju, 500-757, Korea*
*Email: vantam@gmail.com, anhhuy@gmail.com, dchoi@chonnam.ac.kr*

## Abstract

*Context awareness is the most important research area in ubiquitous computing. In particular, for smart home, context awareness attempts to bring the best services to the home habitants. However, the implementation in the real environment is not easy and takes a long time from building the scratch. Thus, to support the implementation in the real smart home, it is necessary to demonstrate that thing can be done in the simulator in which context information can be created by virtual sensors instead of physical sensors. In this paper, we propose ISS, an Interactive Smart home Simulator system aiming at controlling and simulating the behavior of an intelligent house. The developed system aims to provide architects, designers a simulation and useful tool for understanding the interaction between environment, people and the impact of embedded and pervasive technology on in daily life. In this research, the smart house is considered as an environment made up of independent and distributed devices interacting to support user's goals and tasks. Therefore, by using ISS, the developer can realize the relationship among virtual home space, surrounded environment, use and home appliances.*

## 1. Introduction

Ubiquitous computing environments provide access to information and computing resources for users at anytime and anywhere [1, 2]. Since home is the place where everybody lives, smart home aim to provide the best services to home habitants [3, 4]. In this environment, applications must be self-adaptive to the environment within which they operate.

For researchers, it is difficult to work in the real smart home since home appliances are very expensive. Besides, collecting information from sensors, reasoning from known databases, and determining appropriate activities are the main steps for self-adaptive applications. Actually, the main key to those applications is context information. However, there should be various sensors and devices for constructing ubiquitous computing environments for self-adaptive applications. Furthermore it is expensive to construct fully the environment. Furthermore, before installing in the real system, we need home simulation to test, to verify first. Also, the smart home simulator is an ideal place to apply various context awareness approach such as rule based, ontology based, case based reasoning. Therefore, prior to development of self-adaptive applications, it is necessary to demonstrate that it is possible to obtain valid context information from virtual sensors instead of physical sensors.

In this paper, we propose a context aware simulation system called Interactive SmartHome Simulator aka ISS. By using ISS, we automatically collect the context information from a smart home and validate the reactions in ways that fit in with the environment. This is the main design goal of the context aware system. Finally, we can improve the productivity and quality of a smart home realization. This paper is organized as follows: Section 2 presents a brief survey of limitation and problems of related works. Section 3 shows ISS infrastructure. Section 4 described the implementation and how it works via presenting a case study. Related studies, our discussions are given in Sections 5. Section 6 sums up and draws a conclusion.

## 2. Related works

A smart home needs smart electronic appliances but it does not mean they are smart homes itself though they provide intelligent services. To fit each user's request for their suitable services, it is necessary to integrated management system.

There have been lots of works on this research area including the big corporations and research groups. As a result, various ubiquitous computing simulators such as the Ubiquitous Wireless Infrastructure Simulation Environment (Ubiwise) and TATUS and Context Aware Simulation Toolkit (CAST) have been proposed. The Ubiwise Simulator is used to test computation and communication devices. It has three dimensional (3D) models that form a physical environment viewed by users on a desktop computer through two windows [5]. This simulator focuses on device testing, e.g., in aggregating device functions and exploring the integration of handheld devices and Internet service. Thus, this simulator does not consider an adaptive environment. TATUS is built using the Half Life game engine. Therefore, it looks like an assembled simulation game. It constructs a 3D virtual environment, e.g., a meeting scenario. Using this simulator, a user commands a virtual character to perform tasks, such as to sit down. This simulator does not consider device simulation [6]. CAST is a simulator for the test home domain. This simulator uses scenario based approach. It has been proposed as a prototype using Macromedia's Flash MX 2004 [7]. However, using Flash MX [8, 9] does not support users to freely control their environment. JoonSeok Park et al. proposed the design structure for smart home simulator regardless of environment factor as well as interaction aspect [10].

There are still disadvantages of the state-of-the-art research and current home simulation approaches have limitations. Generally, most of the current simulators were not built as the real-time application, and not taken account the real-time aspect in building the system. The related works also have no environment effects and no user interactions in the implementation. Therefore, our approach considers the following important principle of the environment: interaction, placement, retrieve environmental information.

## 3. The proposed simulator overview

The aim of our proposed simulator is to provide the interactive actions in smart home environment and attempt to solve the listed problems. Fig. 1 shows the important components interacting towards the others in general.

- Sensor Retriever: requests and receives sensed information from virtual sensors. There are many kind of sensors in reality can be listed as light, temperature, humidity, location, person, etc.
- Reasoning: queries and concludes the appropriate actions to the current context.
- Smart Home UI: is the tool which helps user to interact with the smart home.
- Home Server: controls the virtual home appliances such as TV, Air conditioner, gate, curtain, electric-based devices like neon light, fan, etc.
- Home Environment: simulates the real environment. Like the real home, home simulator is the container environment consists of smaller environments like living room, bathroom, dining room, etc. Extrinsic factors influence the environment. Exchange of message among extrinsic factors and the environment

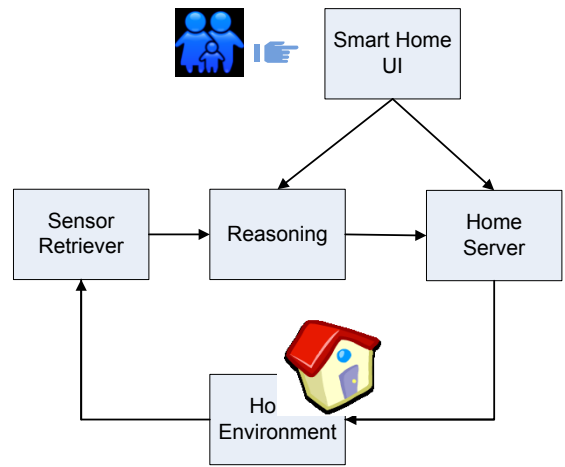

**Fig. 1. The components of proposed framework for smart home simulator**

Those components have relationships to the others. Reasoning part receives data from sensor simulator as well as creates the event and sends to Home Server. Meanwhile, Home Server affects the home environment and Home Environment causes the sensor data to be changed. The home owners, Users, use Smart Home UI to control the Home Server and also affect the reasoning part. Furthermore, other elements also need to be modeled and simulated as follows: Virtual Space, Surrounded environment, Person, Home Appliance.

## 4. Implementation

In this section, first of all, we present our implementation for important components. Then, we introduce some techniques used to make the application more effective.

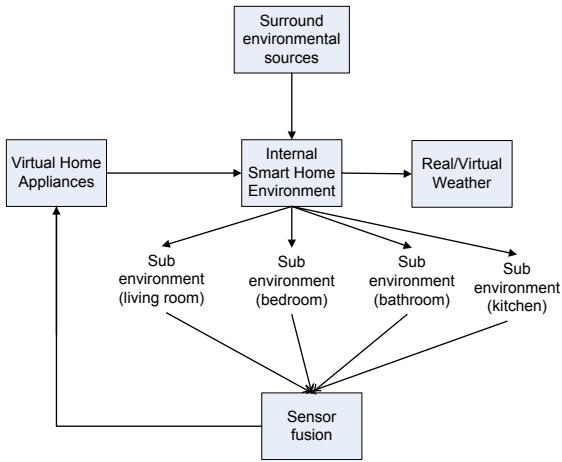

**Fig. 2. Environmental factors in smart home simulator**

### 4.1 ISS components realization

Many researches assume smart home processes one room which means smart home simulator is equal to smart room simulator. However, virtual smart home space consists of many child environments. For example, the internal smart home space includes living room, bedroom, bathroom, kitchen, etc. So, the mission of the simulator is to implement all of those. The simulator also makes the surrounded environment which takes account into the weather element. In fact, we tend to get real data by using Web Service connecting to Yahoo Weather services in order to get the surrounded weather of our area, with the Kwangju region code. But it certainly takes such a long time to wait until the next change of the weather. Therefore, we provide the ability to stimulate the weather to the developers so that they can change the status of the weather belongs to four types of typical sorts of weather in Korea: rain, snow, hot, cloudy. From the weather element, the sensed internal home context information including temperature, humidity, light can be retrieved. When the weather status is updated, it causes the update of the internal house environment information. Fig. 2 illustrates the phase in which the virtual sensor collector receives sensed context information from sensors and transform them into the high-level context.

In order to express the touch and feel in smart home simulator, we consider the implementation of user with user appearance, user movement, user behavior, home appliance element: Placement, Status, and environment as well. The *Factor* component is implemented as the basic atom which is derived by *Person*, *Home Appliance*, *Environment* factor. Virtual Space possesses the list of factors in aggregation relationship. Aggregation differs from ordinary composition in that it does not imply ownership. In composition, when the owning object is destroyed, so are the contained objects, whereas, this is not necessarily true in aggregation. For example, a virtual space consists of many child virtual spaces such as bedroom, living room, etc. And each virtual space owns corresponding person, home appliance and environment factor. If the child virtual space closes, its own factors will continue to exist by moved to the new child virtual space. Therefore, a virtual space can be seen as an aggregation of factors. Additionally, the most important part, the context reasoning part uses rules and case based reasoning. The relationship of those elements is illustrated in Fig. 3. In *the restricted space of this paper* it is impossible to describe explicitly the details of each function in those classes.

### 4.2 The prospective techniques

We use .NET and IDE Sharp Develop [11] which are open sources and free to access. SharpDevelop is an Integrated Development Environment (IDE) for .NET Framework applications. It supports the development of applications written in C#, Visual Basic.NET and Boo. It is open source and written in C#. It provides all of the features required from a modern Windows IDE, such as code completion, project templates, an integrated debugger and a forms designer. Moreover, to make the effect such as the action such as entering or leaving smart home, we use thread and synchronization techniques. Some more techniques can be listed as using socket and message communication; building Sensor simulator; building Smart Home UI; integrating Reasoning part. Those techniques make the simulation more vivid and lively.

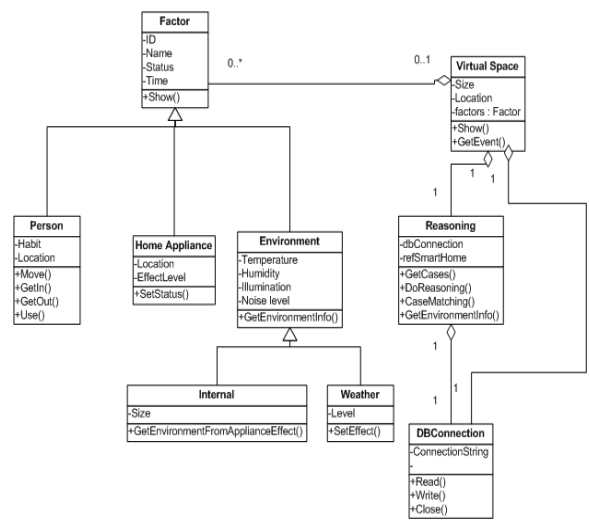

**Fig. 3. The relationship of classes in smart home simulator implementation**

## 5  Achievements

In this section, we present our achieving results and evaluation. After the scenario introduction, we show some achievement results from this.

### 5.2 Case study

We realized the prototype case study as the following:

*"Currently, this is the autumn season, the nicest season in Korea; the temperature is about 20 to 25 Celsius degree. When Mr. Lee arrives home and after certification, he enters home and the light goes on. Also, the air conditioner is turned on and switched to cooling mode. After changing his clothes, he sits on the sofa. At this time, the context manager, detecting his sitting on the sofa, turns on the TV to a channel based on his favorite. When he goes out to do something, every appliance at home will be turned off"*

We chose TV for an example of the application. This case study seems to be very simple but it is practically more complicated in the real situation.

### 5.3 The achievements

To prove that ISS can simulate the usual cases in the real home, we process the above case study on our simulator. Fig 4 shows the results when operating the case study. In Fig. 4a, Mr. Lee is not at home, and everything inside is off. At this moment, the current weather is normal. When Mr. Lee authenticates and enters home, the light is turned on and the AC is also switched to cooling mode, the temperature decreases from 29 to 25 degree Celsius. The simulator updates the information of temperature as well as illumination. When the location sensor detects that Mr. Lee moves to Sofa position, the smart home simulator supposes Mr. Lee uses the sofa and the TV has been set to be turned on and switched to his favorite channel as shown in Fig. 4b.

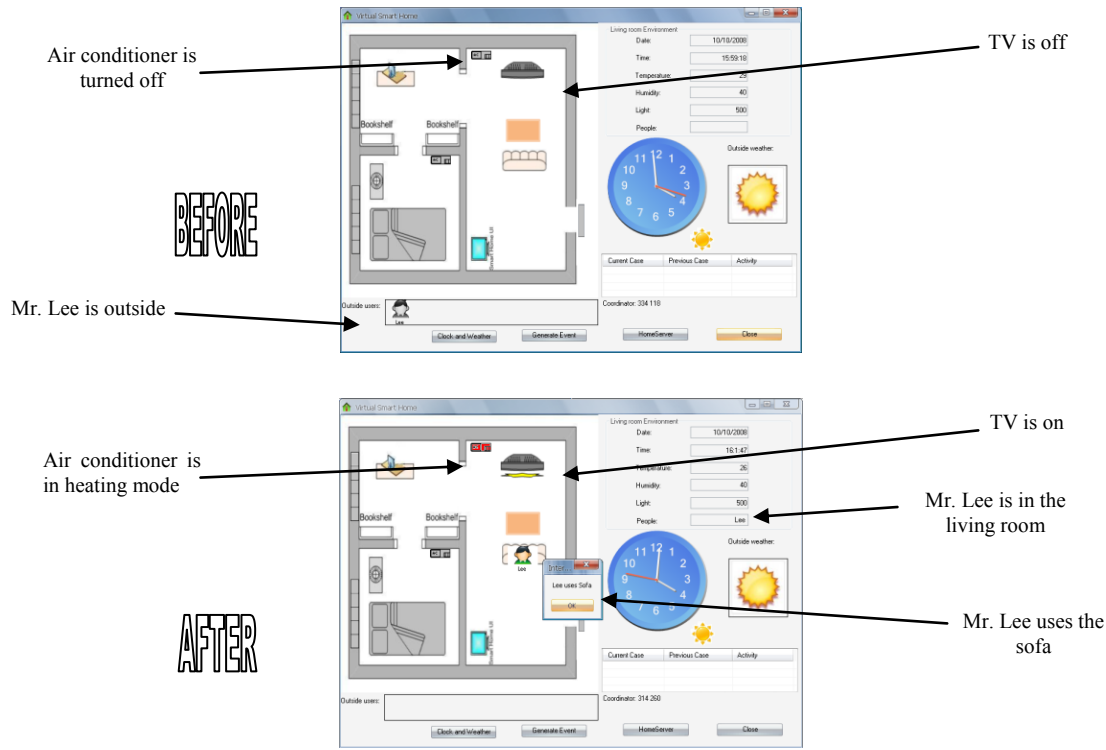

**Fig. 4. The result of before and after states of ISS when operating the case study**

## 6 Discussion

Obviously, making the simulator is always easier than doing the same thing in the real environment. However, working with simulator help us to anticipate what will confront in the near future. Many parts don't look easy to realize as they are in the simulator. For example, in the user location sensor, implemented this part in the real environment have a big trouble, in order to recognize the people, we use RFID tag and RFID reader [12], to recognize the location, we use ZigBee based on RSSI [13]. However, to know "who is there" is the big question. And there are so many other troubles to face when working in the real environment.

In addition, the implementation of Smart Home Simulator which is developed in Java and C# .NET also has another discussion. In common, everyone implies these system boards are so small of weak to install the cumbersome software when mentioning about the middleware for smart home. However, the development of modern board causes the configuration is eligible to set up the latest version of various operating systems such as Windows, Linux, Mac OS. Thus, the simulator can be realized in the real system for both current and upcoming system boards.

## 7 Conclusions

In ubiquitous computing environments, ubiquitous applications are not small and stand-alone but are complex system. In this paper we present the Interactive SmartHome Simulator (ISS) that reflects the relationship between the environment and other factors in smart home. This paper also describes the ISS system architecture and hierarchical rule structure model for Smart Homes. By using this simulator, users can customize the environment at ease such as determining the optimal union sensor and device placement.

In the future, we plan to extend the simulator in order to handle and process complex context information like profile information and meta-data.

## 8 Acknowledgment

This research was supported by the Industry Promotion Project for Regional Innovation. The authors would like to thank the BK21 Project of Korea and the anonymous reviewers for useful comments.

## 9 References

[1] Mark Weiser, "Ubiquitous Computing", Nikkei Electronics, pp.137-143, December 1993

[2] M. Weiser, "the Computer for the 21st Century", American Science, 1991, pp.3-9.

[3] Sei J., "Research Activities on Smart Environment", The Institute of electronics engineers of Korea Journal, pp. 1359-1371, December 2001.

[4] Chemishkian, S., "Building smart services for smart home",Networked Appliances, p.p 215-224, January 2002

[5] John J. B., and Vikram V., "UBIWISE, A Ubiquitous Wireless Infrastructure Simulation Environment", http://www.hpl.hp.com/techreports/2002/HPL-2002-03.html?jumpid=reg_R1002_USEN

[6] O'Neill, E.,Klepal, M.,Lewis, D. O'Donnell, T.,O'Sullivan, D., and Pesch, D., "A testbed for evaluating human interaction with ubiquitous computing environments", Testbeds and Research Infrastructures for the Development of Networks and Communities Conference, pp. 60-69, Feb. 2005.

[7] InSu K., HeeMan P., BongNam N., YoungLok L.,SeungYong L., and HyungHyo L., "Design and Implementation of Context Awareness Simulation Toolkit for Context learning" , IEEE International Conference on Sensor Networks, Ubiquitous, and Trustworthy Computing, vol. 2, pp. 96-103, 2006.

[8] J. Kaye, D. Castillo, "FlashTM MX for Interactive Simulation", ISBN:14-0181-291-0, THOMSON, 2005

[9] Craig Swann, Gregg Caines, "XML in FlashTM",ISBN:0-672-32315-X, QUE Publishing, 2002

[10] JoonSeok Park, Mikyeong Moon, Seongjin Hwang, Keunhyuk Yeom, "CASS: A Context-Aware Simulation System for Smart Home", in Proc. of Fifth International Conference on Software Engineering Research, Management and Applications, 2007

[11] SharpDevelop homepage, http://www.icsharpcode.net/OpenSource/SD/

[12] Roy Want, "An Introduction to RFID Technology", PERVASIVE computing Journal, 2006

[13] G. Ferrari, P. Medagliani, S. Di Piazza, M. Martalò, "Wireless sensor networks: performance analysis in indoor scenarios", EURASIP Journal on Wireless Communications and Networking, Volume 2007 Issue 1, January 2007